# Towards Conceptual Modeling Semantics: Eventizing Tarski's Truth Schema

**Sabah Al-Fedaghi**
*sabah.alfedaghi@ku.edu.kw*
Computer Engineering Department, Kuwait University, Kuwait

**Summary**
Modeling languages in software engineering (e.g., UML) evolved from software systems modeling where denotational and operational kinds of semantics are the traditional subjects of research and practice. According to some authors, although a large portion of the static semantics (e.g., UML) seems to have reached a consensus, the dynamic semantics of activities, interactions, and state machines poses a major challenge. Central to semantics is the relationship between a sentence and the (actual) world. Carefully examining semantics-related issues in the modeling languages field to avoid problems that may affect practical applicability is important. One effort in this direction is OMG's release of a 2020 draft specification for Foundational UML (fUML), with the base semantics specifying executions that are executable in the same sense as a program in a traditional programming language. Additionally, efforts within academia have sought to develop an alternative approach to modeling languages using formal semantics (e.g., using Russell's theory of types and Tarski's declarative semantics). This paper aims at a similar exploratory venture of developing semantics, only for a much more modest diagrammatic modeling language, called the thinging machine model. The model promotes a deep understanding of the scrutinized modeling language and leads to considerably fruitful questions. Constructing the thinging machine model seems to facilitate progress in this direction, and the initial results in this paper indicate the viability of the approach.
*Key words:*
*Conceptual modeling*, *modeling language*, *diagrammatic representation*, *formal semantics*, *UML semantics*, *T-schema*

## 1. Introduction

Diagrammatic modeling languages (e.g., UML) have become the key artifacts of software development, and "the solidness of modeling languages metamodels" has generated important outcomes in the field of software engineering [1]. They are proving extremely helpful in software and systems development. However, when formal semantics (meaning in contrast to syntax) are called for, properly defining diagrams seems to be a much harder problem [2]. In the case of UML, the language evolved from software systems modeling where denotational (using mathematical objects) and operational (using execution) kinds of semantics are the traditional subjects of research and practice [3-4]. According to Ekenberg and Johannesson [5], work on formalizing UML has attempted to use different versions of temporal logic (see, e.g. [6-7]). Dynamic logic has also been used as a basis for UML semantics (e.g., in [8-9]). In 2020, OMG [10-11] released a draft specification of Foundational UML (fUML) wherein basic semantics specify when particular executions conforming to a model defined in fUML generate executions. As a specification, fUML has standard, precise execution semantics. It is a subset that includes constructs of UML and the ability to model behavior using a composed set of primitive actions. A model constructed in fUML is "executable in exactly the same sense as a program in a traditional programming language, but it is written with the level of abstraction and richness of expression of a modeling language" [12].

According to Broy et al. [13], although a large portion of the static semantics of UML seem to have reached a consensus, the dynamic semantics of the UML sublanguages, such as activities, interactions, and state machines, pose a major challenge. These foundational difficulties involving dynamic semantics of such tools as state machines lead to definitions that contradict common sense [13]. They "show how important it is to carefully design a modeling language to avoid problems regarding its expressivity as well as its interpretation—problems that strongly impact practical applicability [13]. Efforts within academia have sought to develop an alternative approach to UML's formal semantics (e.g., using Russell's theory of types and Tarski's declarative semantics [1-4]).

1.1 Aim
This paper aims at a first step toward developing semantics, only for a much more modest diagrammatic modeling language called the thinging machine (TM). According to Broy et al. [13], formalization is a scientific approach that promotes deep understanding of different aspects of the scrutinized modeling language. Trying to conduct the formalization uncovers many properties of the modeling language and leads to considerably fruitful questions. In this paper, we explore a new territory in logic as much as in modeling though investigation of the notion of *truth* and hence the semantics for relations among the concepts and the subject that is being modeled in the TM context.

### 1.2 Justification

The paper suggests applying logical semantics to the diagrammatic method used in conceptual modeling. The justification for this direction of research is that the reliance on graphical constructs poses problems when it comes to precise and unambiguous semantics [5]. The first-order logic, when used in conjunction with conceptual modeling, "provides a sound basis on which specifications written in a process-based language can be transformed, merged, and verified for the purpose of detecting interference" [5]. Additionally, logic-based formalization provides support for tasks by means of logic-based inference (a topic not discussed in this paper) so that tools can provide effective support for automated reasoning [14].

Diagramming has been used in logic for various reasons. Besides the use of diagrams as illustrations or thought aids, diagrammatic systems have also formalized logic. These include Frege's [15] Begriffsschrift ("conceptual notation") and Charles Sanders Peirce's [16] existential graphs.

### 1.3 Outline

TM modeling is a three-level process that involves the following:
- A static model of the state of affairs to produce an atemporal diagrammatic description denoted as S.
- A decomposition of S into subdiagrams that form the base of temporal events.
- The behavior of the model, denoted as B, is formulated as a chronology of events. The behavior refers to executing composite actions.

We try to clarify the intuitive conception of the TM diagrammatic representation by incorporating the semantics notion of logical consequence. The fundamental idea is that the semantics are built upon B (chronology of permitted events).

Thus, consider a given statement p, in terms of Tarski's famous T(ruth) schema [17]:

"p" is true if and only if P. (i.e., iff the corresponding state of affairs holds)

The T schema can be formalized in many-sorted predicate logic or modal logic. Tarski conceived the T schema as an expression of the classical correspondence theory. This conception was done in linguistic terms that are supposed to refer to objects in the world. Our basic idea is to "inject" the diagrammatic form as follows:

"p" is true if, and only if, B (i.e., iff the corresponding chronology of events holds)

where B is the chronology of events expressed as a diagrammatic construct. This form is generalized for S, and hence, if S is true, then so is B. Although such a form of representation does not bring a new idea to Tarski's T schema, it weakens its reliance on textual language because B is specified as a diagrammatic expression.

To achieve a self-contained paper, the next section reviews the TM model. A more elaborate discussion of the TM model's foundations can be found in [18-29].

## 2. The TM Model

The main TM thesis is that each entity has a double nature as (i) a thing and (ii) a process (abstract machine); thus, we call these thing/machine entities *thimacs*. In TM modeling, intertwining with the world is accomplished by integrating these two modes of being of entities. Thimacs inhibit the traditional categorization, properties, and behavior, replacing them with creating, processing, releasing, transferring, and receiving. Such a thesis has profound influence on the semantics of TM modeling of the world. It implies that all actions are reduced to five actions or generic (elementary) machines. Because machines are things, all things can be reduced to five elementary things: the create thing, the process thing, the release thing, the transfer thing, and the receive thing. These ideas were inspired by and can be traced back to Aristotle in ancient history and Heidegger in modern times (see [29]). As stated in Al-Fedaghi [29], Aristotle proclaimed entities are the sorts of "basic beings that fall below the level of truth-makers, or facts, just as … nouns and verbs, things said 'without combination,' contribute to the truth-evaluability of simple assertions" [30]. Moreover, Aristotle introduced the notion of process in thinking about things. He conjectured that a thing in nature persists via an internal process that must be realized within a matter that harbors tendencies resulting from its elemental components (e.g., fire, water, earth, or air). This causes tendencies to actively strive toward their "natural place." In this view, Aristotle can be counted as a process philosopher [31]. Aristotle's idea is that things are compounds consisting of matter and form. According to Heidegger [32], a thing is self-sustained, self-supporting, or independent—something that stands on its own. The condition of being self-supporting transpires by means of producing the thing. Heidegger [32] encourages further research on "generic processes" applied to a thing.

Accordingly, in this paper, we claim that mapping to the "object world" can be accomplished by mapping the five generic thimacs. In TM modeling, a thing's machine operates on other things by creating, processing, releasing, transferring, and/or receiving them. The term "machine" refers to a special abstract machine (see Fig. 1). The TM description of a system is built under the postulation that it only performs five generic operations: creating, processing (changing), releasing, transferring, and receiving. A thing is created, processed, released, transferred, and/or received. A machine creates, processes, releases, transfers, and/or receives things. Among the five stages, flow (a solid arrow in Fig. 1) signifies conceptual movement from one machine to another or among a machine's stages.

The TM's actions (called also stages) can be described as follows:
- *Arrival*: A thing reaches a new machine.
- *Acceptance*: A thing is permitted to enter the machine. If arriving things are always accepted, then arrival and acceptance can be combined into the "receive" stage. For simplicity, this paper's examples assume a receive stage exists.
- *Processing* (change): A thing undergoes a transformation that changes it without creating a new thing.
- *Release*: A thing is marked as ready to be transferred outside of the machine.
- *Transference*: A thing is transported somewhere outside of the machine.
- *Creation*: A new thing is born (is created/emerges) within a machine. A machine creates in the sense that it finds or originates a thing; it brings a thing into the system and then becomes aware of it. Creation can designate "bringing into existence" in the system because what exists is what is found. Additionally, creation does not necessarily mean existence in the sense of being alive. Creation in a TM also means appearance in the system. Appearance here is not limited to form or solidity but also extends to any sense of the system's awareness of the new thing.

In addition, the TM model includes
- Memory
- Triggering (represented as dashed arrows), or relations among the processes' stages (machines); for example, the process in Fig. 1 triggers the creation of a new thing.

To approach TM modeling smoothly, we focus on the machine side of thimacs. The duality of a thimac will be examined later in the paper.

## 3. TM Modeling

Klimek [33] dealt with the problem of the lack of tools for automatic extraction of logical specifications from software models and proposed a method for automatic generation of these specifications, considered as sets of

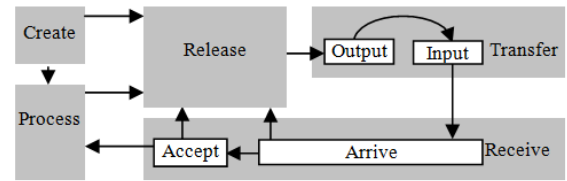

Fig. 1  Thinging machine.

temporal logic formulas. Klimek [33] illustrated the approach by considering a "business use case scenario" to illustrate behavior, where the scenario allows for identification and extraction of atomic activities. This is followed by developing a UML activity diagram to enable the modeling of atomic activities. A sample scenario given by Klimek [33] is as follows:
1. Passenger's "Check-in" or "selfCheck-in"
2. If necessary, then "HoldBaggage"
3. If non-Schengen, then "BoarderControl" and "CustomControl"
4. Passenger's "securityControl"
5. Passenger's "Board"

Klimek [33] also used "use case diagrams" to model this scenario. Propositions (atomic activities) were declared, such as

$Seq(Seq(Branch(a, b, c), Branch(d, e, n1)))$ and
$Seq(Branch(f, Seq(g, h), n2), Seq(i, j))$,

where $a$ is Counter, $b$ is CheckIn, $c$ is SelfCheckIn, $d$ is Baggage, $e$ is HoldBaggage, and so on. Accordingly, a logical specification is developed. For example, e ⇔ j means that if the HoldBaggage for a passenger is registered, then sometime in the future the passenger will board—or, more formally, *HoldBaggage* ⇒ *Boarding*.

### 3.1 Static TM Model

Fig. 2 shows the static TM model, S, developed according to our understanding of the scenario. The figure describes two types of passengers (circle 1): with luggage (2) and without luggage (3). The passenger with luggage moves (4) to the counter, where his or her luggage is received (5) and processed (6). At the counter, the passenger is processed to be given a travel ticket (7) and moves to the queue area (8). The passenger without luggage goes to the self-service check-in (9), is processed (10), and moves to the queue area (11).

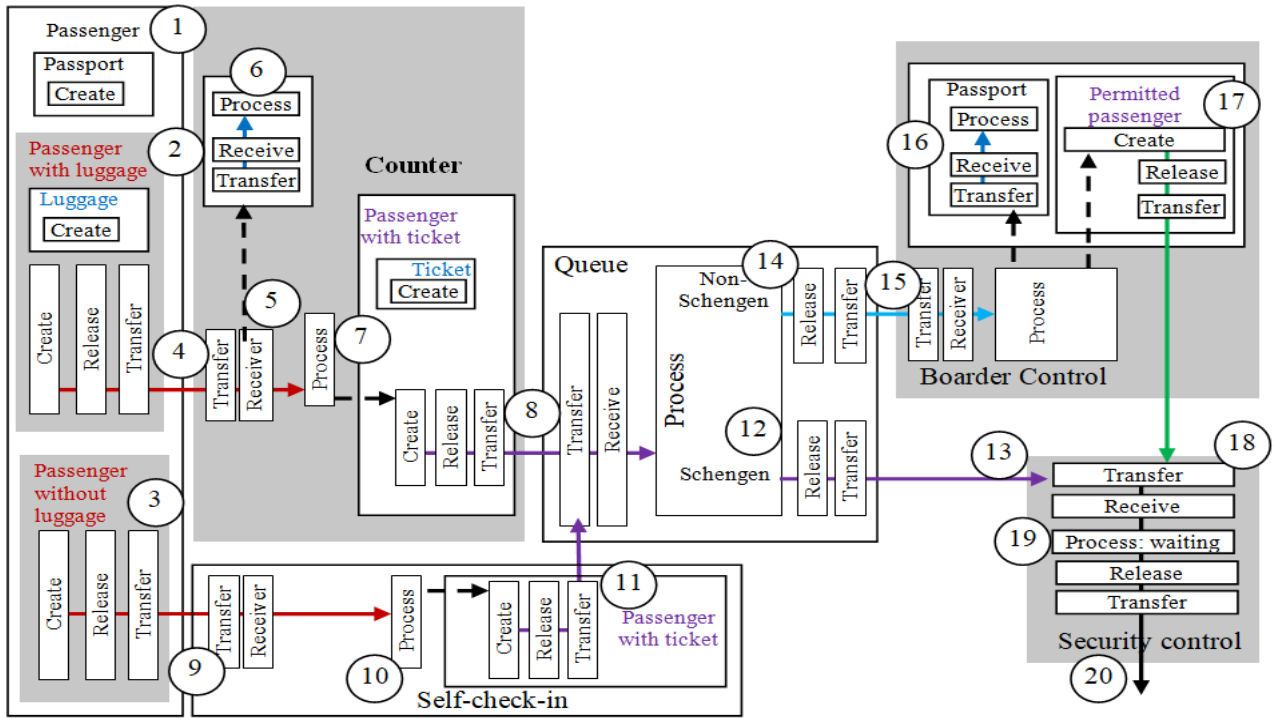

Fig. 2. The static TM model of the airport passenger scenario.

In the queue area, if the passenger is of type Schengen (12), he or she proceeds to the security control area (13); if not of that type (14), then the passenger goes to the boarder control area (15), where his or her passport is processed (16). Assuming that everything is acceptable, the passenger is permitted to move to the security control area (17 and 18) and waits there (19) until boarding (20).

### 3.2 Decomposition of the Static Model

In Fig. 2, S is a static description that represents all states of affairs. A *state of affairs* is a combination or complex of thimacs. We need a "structure" for this complex to reduce it to a multiplicity of "meaningfulness." Fig. 2 is reminiscent of Deleuze and Guattari's philosophical notion of a "body without organs" [34]. S is a "body" that has the potentialities of phenomena, as an airport mechanism that can be populated by organs (e.g., handling luggage, boarder control, or self-service ticketing) or subsystems, each with its own purpose. S is also a source of the system behavior to be, if we can figure out how to make it into an assemblage of organs that form a goal-directed organization as a thimac: a thing and a machine.

This discussion points to S as a machine schema that is amenable to compositional exploration to generate a new structural level (multiplicity). The *structure* of a particular composite of unity is the manner in which it is made by actual static components in a particular space as well as a particular composite unity. The point of this discussion is to view S as an organization that needs structuring so that its behavior can be specified. While the wholeness of S is the same, S may have different structures depending on how it is divided into parts.

The idea of decomposing a system for semantics analysis is taken from the study of semantics in languages. In so-called compositional semantics, the truth value of a sentence is calculated by composing, or putting together, the meanings of smaller units [35]. The meaning of a statement is composed of the meanings of its parts and how they are combined structurally [35].

### 3.3 Subdiagrams (Changes) in S

Fig. 3 shows the decomposition of the diagram S into 14 subdiagrams: $S_1, S_2, \ldots S_{14}$. These subdiagrams replace S with potential locations of changes. A change in the S model refers to *becoming* different or *becoming* altered or modified. Each subdiagram is assigned a name as follows:

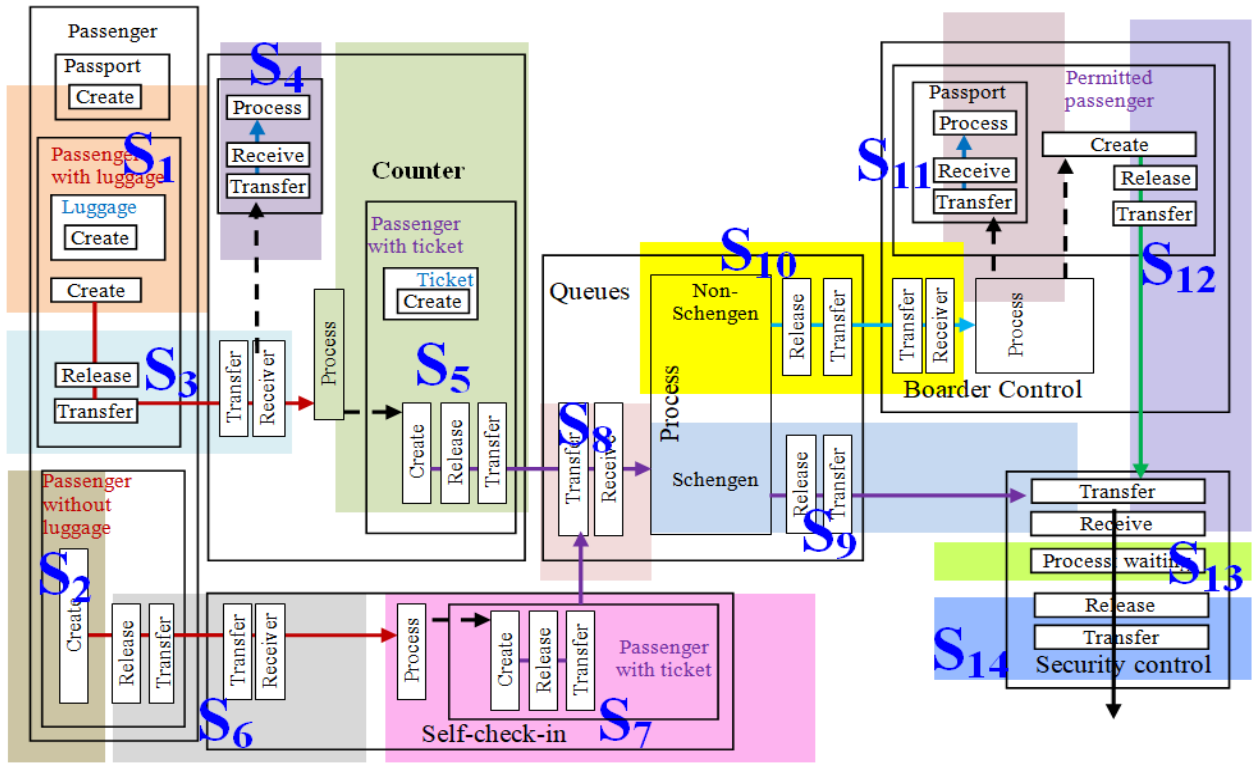

Fig. 3 The TM dynamic model of airport passenger processing.

Note that these names are written in a certain style to emphasize that they are subdiagrams and not language strings.
$S_1$: PASSENGER-WITH-LUGGAGE-IS-PRESENT
$S_2$: PASSENGER-WITHOUT-LUGGAGE-IS-PRESENT
$S_3$: PASSENGER-WITH-LUGGAGE-MOVES-TO-THE-COUNTER
$S_4$: LUGGAGE-IS-RECEIVED-AND-PROCESSED-AT-THE-COUNTER
$S_5$: PASSENGER-WITH-LUGGAGE-IS-PROCESSED-TO-BE-A-PASSENGER-WITH-TICKET-AND-LEAVES-THE-COUNTER
$S_6$: PASSENGER-WITHOUT-LUGGAGE-MOVES-TO-THE-SELF-SERVICE-AREA
$S_7$: PASSENGER-WITHOUT-LUGGAGE-IS-PROCESSED-TO-BE-A-PASSENGER-WITH-TICKET-AND-LEAVES-THE-SELF-SERVICE-AREA
$S_8$: PASSENGER-WITH-A-TICKET-ARRIVES-AT-THE-QUEUE-AREA
$S_9$: PASSENGER-WITH-A-TICKET-IS-PROCESSED-AT-THE-QUEUE-AREA-AND-IDENTIFIED-AS-A-SCHENGEN-TYPE-AND-MOVES-TO-THE-SECURITY-CONTROL-AREA
$S_{10}$: PASSENGER-WITH-A-TICKET-IS-PROCESSED-AT-THE-QUEUE-AREA-IS-IDENTIFIED-AS-A-NON-SCHENGEN-TYPE-AND-MOVES-TO-THE-BOARDER-CONTROL-AREA
$S_{11}$: AT-THE-BOARDER-CONTROL-AREA-THE-PASSENGER-HAS-HIS/HER-PASSPORT-PROCESSED
$S_{12}$: AT-THE-BOARDER-CONTROL-AREA-THE-PASSENGER-MOVES-TO-THE-SECURITY-CONTROL-AREA
$S_{13}$: PASSENGER-WAITS-FOR-BOARDING-AT-THE-BOARDER-CONTROL-AREA
$S_{14}$: PASSENGER-LEAVES-THE-BOARDER-CONTROL-AREA-TO-BOARD-THE-PLANE

## 4. Behavioral Model, B

Eventually, this decomposition aims to reconceive S in terms of events: actual existent things (thimacs) that form the semantics of S. In parallel with Tarski's T schema condition [36], "It rains" is true iff IT RAINS; we will declare that $S_i$ is true iff it is eventized, $1 \leq i \leq 14$.

An event in the TM model is defined as a thimac with a time subthimac, which is a subdiagram of S with a time machine. For example, Fig. 4 shows the event *A passenger with a ticket is processed at the queue area, is identified as a non-Schengen type, and moves to the boarder control area*. Note that the subdiagram in this event is $S_{10}$.

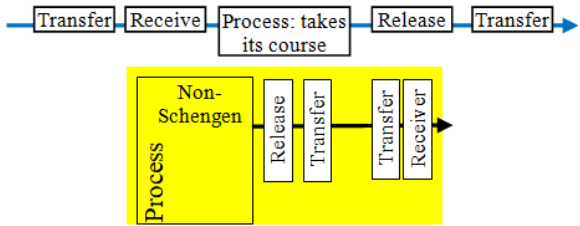

Fig. 4. The event *A passenger with a ticket is processed at the queue area, is identified as a non-Schengen type, and moves to the boarder control area.*

The event may have another submachine—say, intensity—but such is not relevant to this discussion. In an analogy to Tarski's condition mentioned above, associating time with a subdiagram amounts to associating the time NOW with IT IS RAIN*ING*.

An event is a period of time in which a thimac materializes. We have projected the thimac materialization in terms of its subthimacs (subdiagrams). Accordingly, we can convert $S_1, S_2, \ldots S_{14}$ to events $E_1, E_2, \ldots E_{14}$ (see Fig. 5, where each event is represented by its subdiagram).

Event 1 ($E_1$): A passenger with luggage is present.
Event 2 ($E_2$): A passenger without luggage is present.
Event 3 ($E_3$): A passenger with luggage moves to the counter.
Event 4 ($E_4$): The luggage is received and processed at the counter.
Event 5 ($E_5$): A passenger with luggage is processed to be a passenger with a ticket and leaves the counter.
Event 6 ($E_6$): A passenger without luggage moves to the self-service area.
Event 7 ($E_7$): A passenger without luggage is processed to be a passenger with a ticket and leaves the self-service area.
Event 8 ($E_8$): A passenger with a ticket arrives at the queue area.
Event 9 ($E_9$): A passenger with a ticket is processed at the queue area, is found to be a Schengen type, and moves to the security control area.
Event 10 ($E_{10}$): A passenger with a ticket is processed at the queue area, is found to be a non-Schengen type, and moves to the boarder control area.
Event 11 ($E_{11}$): At the boarder control area, the passenger has his/her passport processed.
Event 12 ($E_{12}$): From the boarder control area, the passenger moves to the security control area.
Event 13 ($E_{13}$): The passenger waits for boarding at the boarder control area.
Event 14 ($E_{14}$): The passenger leaves the boarder control area to board the plane.

It is not difficult to write these events in terms of propositional functions. For example, *A passenger with luggage is present* can be written as *is-Present (x)*, where the domain of *x* is passengers with luggage, and *A passenger with luggage moves to the counter* can be written as *moves (x, y)*, where *y* is the counter.

We can claim the following:
$S_i$ is true iff $E_i$, $1 \leq i \geq 14$.

$\{E1, E2, \ldots E14\}$ has a chronology of events, as shown in Fig. 5, that expresses the behavior B of the system. In general, we can conclude that

S is true iff B.

Here, the word *true* expresses a property of diagrams. The diagrammatic language contains the capacity to refer to its own subdiagrams (expressions), and thus the events language can be considered the meta-language of the object diagrammatic language that expresses S.

## 5. Behavioral Definitions of Action

TM modeling is based on thimacs (things/machines), which is denoted by Δ. Δ has a dual mode of being: the machine side, denoted as M, and the thing side, denoted by T. Thus, Δ = (M, T).

Fig. 6 shows the generic action in the T machine. In the context of semantics, these actions are words of sentences in the study of language semantics. The semantics are analogous to so-called lexical semantics (word meaning). In this section, we present a preliminary attempt to bound semantics to five events of the T's five generic actions. Since Δ = (M. T) under the duality assumption, the five generic events apply to things.

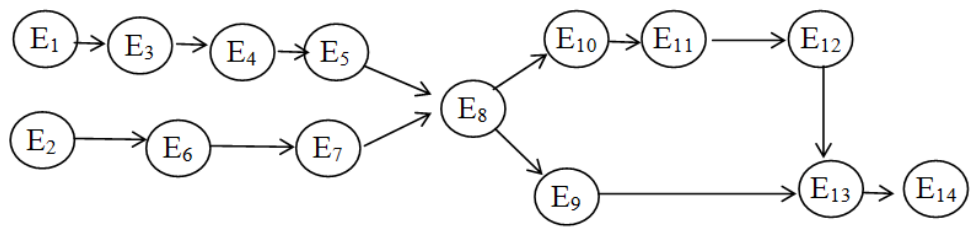

Fig. 5. The TM behavioral model of airport passenger processing.

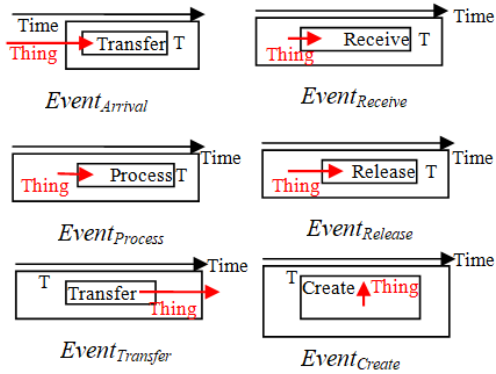

Fig. 6 The semantics of different generic actions.

We can now give a behavioral definition of the actions in the machine of Fig. 1.

*Arrival* ≡ (is defined as) *Event<sub>Arrival</sub>*; that is, the meaning of arrival (i.e., a machine with only the state of arrive) is the event of a thing entering the boundary of any machine. Thus, Arrive is true iff *Event<sub>Arrival</sub>*. For example, "John arrives to London" is true iff the event *John is moving from the outside to, say, inside the perimeter of London* occurs. Such an action is not (to our knowledge) recognized in different bodies of literature. In set theory, $x \in A$ means that $x$ is/has become a member of A. On the other hand, *Arrive* "means" "depositing" $x$ in the set, yet it does not become a member until it reaches *Receive*. This event is represented in Fig. 6, where, for simplicity's sake, the stages of the time are deleted.

The semantics of each action can be defined in a similar way, as shown in Fig. 6. Each of the generic actions in Fig. 6 represents a generic event. The event is specified by a single TM action and time. Accordingly, the semantics of events with larger TM subdiagrams can be mapped to these generic events. Because a machine is a thing, the machines in Fig. 6 are also "events of things" as much as they are events of actions. Each machine is an event thimac. When time is removed, each subdiagram represents a thimac.

Consider the single-stage thimac Create, which only creates and does nothing else. Let us denote this thimac with Δcr. In this Δcr monistic world, Δcr events generate only Δcrs. Δcr does not process, release, transfer, or receive and is similar to Leibniz's monads because it is "simple," having no parts and therefore being indivisible. It may have memory.

Similar accounts can be presented for other generic thimacs in TM modeling that we will not elaborate on in this paper. The point here is that these primitive things/machines are the nuclei of primitive (do not embed subthimacs) behaviors. The informal meanings associated with them are as follows:
- Existing/appearing (create).
- Crossing a boundary (transfer)
- Becoming an element (receive)
- Changing in form (process)
- Dismissing membership (release)

Such an initial treatment of basic semantics needs more formal treatment, but the method to accomplish that is clear.

## 6. Semantic Events in Linguistics

A related topic to this paper is semantic events introduced by Davidson [37], where events are viewed as spatiotemporal things (i.e., concrete particulars with a location in space and time). Consider the sentence from [38] *Jones buttered the toast* and its logical form: *Butter (jones, the toast)*. According to Maienborn [38], "Davidson (1967) points out such a representation does not allow us to refer explicitly to the action described by the sentence and specify it further by adding, e.g., that Jones did it slowly, deliberately, with a knife, in the bathroom, at midnight." According to Davidson, action verbs introduce an additional hidden event argument that stands for the action proper. Davidson proposed expressing the above statement with ∃e[*Butter (jones, the toast, e)*].

Clearly, the topic of semantic events in linguistics is related to the events in the conceptual model. This issue needs further exploration in future research. For the time being, we will not try to mix the issue of events in these two approaches. However, in anticipation of such a development, we explore next some samples of modeling sentences in TM.

## 7. Applying the Method

The sphere of interest in this section of the paper is limited to linguistic expression (including logic language). We examine a number of linguistic expressions as carriers of meaning. In this method, "understanding" a statement begins with translation of it into a TM diagram. This translation may resolve ambiguities and incorporate implicit information. The static diagram is decomposed, and events are identified to construct the corresponding TM diagram that represents the behavioral TM model B. Accordingly, the statement is true iff B.

In this process, the linguistic expression is translated in a more suitable language for semantics. The TM language has five generic actions, and thus the nuclei of meanings are limited. Second, the totality of the description is decomposed into "meaningful" pieces. The limits of the pieces are the actions/things: create, process, release, transfer, and receive. Generic meanings are then connected with time in terms of events, thus causing further confinements of meaning. Additionally, the chronology of events restricts the involved interpretations.

Note that we will use a simplified version of TM modeling whereby the actions release, transfer, and receive are eliminated under the assumption that the direction of the arrow in the diagram is sufficient to indicate the flow of things.

### 7.1 Example: *The moon is made of green cheese*

A proposition is a declarative sentence that is either true or false. In TM, a proposition is a machine that has a submachine called a *truth value*. Consider the proposition *The moon is made of green cheese* (or "The Moon is made of green cheese.") as a thimac. Its machine representation is shown in Fig. 7. In the figure, the proposition (1) has three components: the English text (2), the TM diagram (3), and the truth value (4).

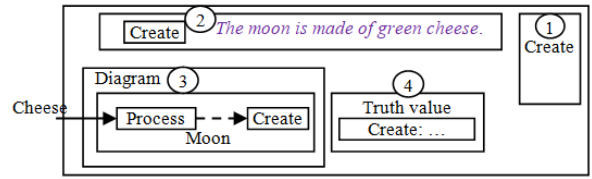
Fig. 7. The proposition *The moon is made of green cheese.*

The figure illustrates the propositional truth assignment according to the correspondence theory of truth. We apply decomposition to the diagram of the proposition, as we did in the previous section, to produce the following events (see Fig. 8):
- $E_1$: Processing cheese
- $E_2$: Creating moon

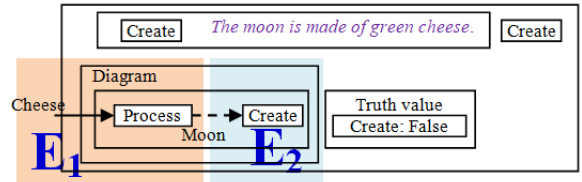
Fig. 8. The events in the proposition.

These events have an order, and $E_1$ is "before" $E_2$. Accordingly, Fig. 9 shows the behavioral model B of the proposition according to the chronology of events. Hence, *the moon is made of green cheese* is true iff B. That is, the proposition is true iff the events $E_1$ and $E_2$ occur.

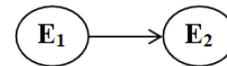
Fig. 9. The behavioral model of the proposition.

### 7.2 Example: *Bread is made of flour and water*

Consider the proposition expressed in the form "Bread is made of flour and water," as represented in Fig. 10 (simplified diagrammatic version) and with the events shown in Fig. 11. According to Tarski's T schema [36], "Bread is made of flour and water" is true iff BREAD IS MADE OF FLOUR AND WATER.

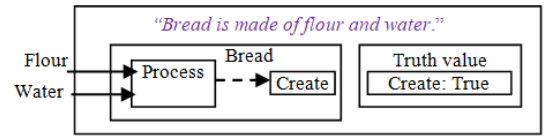
Fig. 10. S of the proposition *Bread is made of flour and water.*

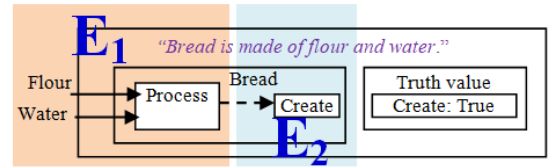
Fig. 11. The events in the proposition *Bread is made of flour and water.*

Similar to the previous example, the diagram in Fig. 10 is true iff it is B. The new thing in this formulation is expressing the original problem in a diagramming language. The TM modeling extends the semantics of S (Fig. 10) to produce the behavioral model B.

"Bread is made of flour and water" is true if and only if B. B is a chronology of events (not shown since it is similar to Fig. 9). Thus, B refers to an event where flour and water are mixed, followed by the event of bread being generated.

### 7.3 Example: *0+0=1*

Consider the proposition *0+0=1*. Figs. 12–14 (simplified version) show the corresponding three diagrams of S, events, and behavioral diagrams. Hence, *0+0=1* is true iff the chronology of events in the behavioral model occurs; i.e., zero is generated twice, the two zeroes are summed, and the summation produces 1.

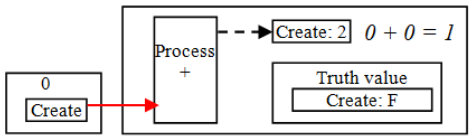

Fig. 12. The S model of *0+0=1*.

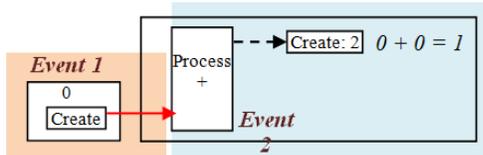

Fig. 13. The events of *0+0=1*.

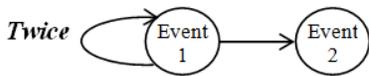

Fig. 14. The behavioral model of *0+0=1*.

### 7.4 Example: *John gave Mary an apple*

According to Nouwen [39], two sentences that entail one another have the same semantic meaning. For instance, *John gave Mary an apple* both entails and is entailed by *John gave an apple to Mary*. This suggests that the dative alternation in English has no semantic import. Figs. 15 and 16 shows the TM static and behavioral models. The truth value of the diagram is assigned according to the behavioral model; that is, it *happens* that John releases and transfers an apple that is received by Mary. Fig. 17 shows the event *John is giving Mary an apple*, which is true if the event is happening now.

### 7.5 Example: *The boy saw the man with the telescope*

Consider the statement (from [35]) *The boy saw the man with the telescope*. Fig. 18 shows two possible TM representations of the statement. In the figure, "create" indicates "there is." In the upper diagram, the man exhibits his image that the boy is using a telescope to see. In the bottom diagram, the image of a man with the telescope is seen by the boy. Clearly, the diagrams when converted to events—say, $B_1$ and $B_2$—are different; thus, ambiguity is eliminated, and the truth depends on which behavior is adopted. We can say that $B_i$, I = 1 or 2 is the *referent* of the given statement.

## 8. The Liar Paradox

*Self-reference* denotes a statement that refers to itself. The most famous example of a self-referential sentence is the liar sentence: *This sentence is false*. The involved paradox that is reflected in such a statement seems to show that truth and falsity actually lead to a contradiction if we apply the following:

*This sentence is false* is true iff THIS SENTENCE IS TRUE.

If the statement is true, then *This statement is false* is true. Therefore, it must be false. If the statement is false, then *This statement is false* is false and therefore must be true.

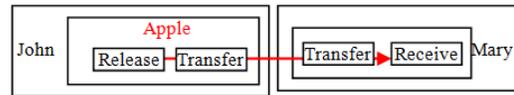

Fig. 15. The S model of *John gave Mary an apple*.

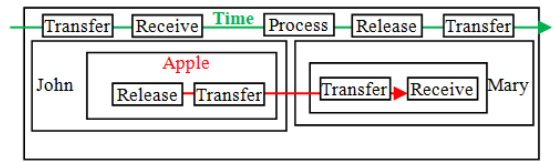

Fig. 16. The event *John gave Mary an apple*.

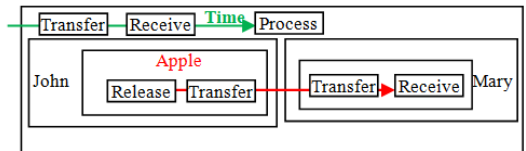

Fig. 17. The event *John is giving Mary an apple*.

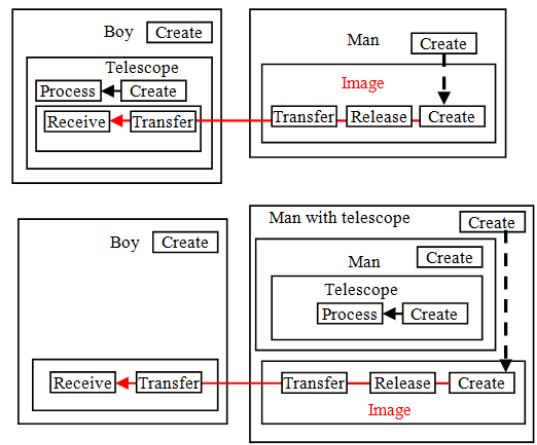

Fig. 18. Two interpretations of *The boy saw the man with the telescope*.

It has been proposed that the statement is neither true nor false and that it is both true and false. In both cases, we end up with infinite regress because the involved statement is self-referential. Alfred Tarski suggested that the paradox arises only in languages and that solving it requires utilizing levels of languages.

In TM modeling, we utilize the notion of chronology of events to eliminate infinite regress resulting from self-reference. Without loss of generality, we use the version of the liar paradox *I am lying*.

As a result of the TM representation, if *I am lying* is true, then it stays true without infinite regress. If *I am lying* is false, then it stays false without infinite regress. *I am lying* can be true or false. It is false if it is used in a sarcastic way where the speaker is saying the opposite of what they really mean, as in saying "early" to mean "late" or "knowledgeable" to mean "ignorant." Alternatively, *I am lying* may be true if it is used in the usual way.

Fig. 19 shows the TM model of the proposition *I am lying*, which involves the following:
1. There is I (create).
2. I process myself (e.g., dwell/practice/activate).
3. I create and process lies.

Hence, we translate "I am lying" into a diagrammatic representation that expresses the existence of the I who creates lies. Fig. 20 shows the events in S. The time sense of NOW in "lying" is a complex event that includes sub-events, just as saying *I am writing* implies I am in the middle of a time period where I am producing consecutive letters, words, and sentences. Similarly, *I am lying* indicates (see the behavioral model B in Fig. 21)

- Event 1 ($E_1$): There is I (create),
- Event 2 ($E_2$): I process myself (e.g., dwell/ practice/ activate), and
- Event 3 ($E_3$): I create and process lies,

in that order. Accordingly, *I am lying* is true iff $E_1 \rightarrow E_2 \rightarrow E_3$, or *I am lying* is true iff B.

We observe that these semantics preserve whatever truth value we assign to the proposition and eliminate infinite regress. Simply, *I am lying* is true iff *there is I* (create-I event) and this I triggers (dash arrow from create to process) creating lies.

## 9. Conclusion

This paper has introduced a first-step venture into developing semantics for the diagrammatic modeling language TM. We explored a new territory in logic as much as in modeling though investigation of the notion of *truth* and hence of the semantics for relations between the concepts and the subject being modeled in TM. This seems to facilitate an unconventional direction that is still in need of scrutiny, but the initial results of this paper indicate the viability of the approach.

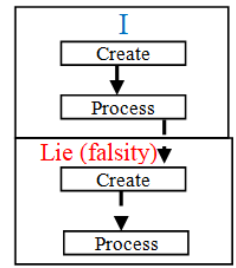

Fig. 19 The diagrammatic model S of *I am lying*.

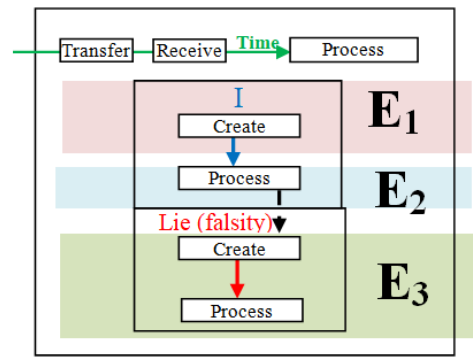

Fig. 20 The events in *I am lying* thimac.

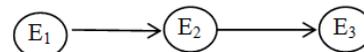

Fig. 21 The behavioral model.

## References

[1] A. Naumenko, "Triune Continuum Paradigm: a paradigm for general system modeling and its applications for UML and RM-ODP," PhD thesis 2581, Swiss Federal Institute of Technology of Lausanne, June 2002.

[2] D. Harel and B. Rumpe, "Meaningful modeling: what's the semantics of 'semantics'?", Computer, vol.37, pp.64-72, 2004.

[3] A. Naumenko, A. Wegmann and C. Atkinson. "The role of Tarski's declarative semantics in the design of modeling languages," Technical report No. IC/2003/43, Swiss Federal Institute of Technology of Lausanne, April 2003.

[4] A. Naumenko and A. Wegmann, "Triune continuum paradigm and problems of UML semantics," Technical Report IC/2003/44, Swiss Federal Institute of Technology of Lausanne, 2003, http://www.triunecontinuum.com/documents/tr03_044.pdf

[5] L. Ekenberg and P. Johannesson, "UML as a first order transition logic," The European-Japanese Conference on Information Modelling and Knowledge Bases, Krippen, Swiss Saxony, Germany, 2002.

[6] A. Knapp, "A formal semantics for UML interactions," in Proceedings of UML'99, eds. R. France and B. Rumpe, LNCS 1723, Springer, 1999.
[7] S. Kim and D. Carrington, "Formalising the UML class diagram using object-Z," in Proceedings of UML'99, eds. R. France and B. Rumpe, LNCS 1723, Springer, pp.Page-Page, Location, 1999.
[8] C. Pons, G. Baum and M. Felder, "Foundations of object-oriented modeling notations in a dynamic logic framework," in Foundations of Models and Languages for Data and Objects, Kluwer, pp.Page-Page, Location, 1999.
[9] G. Övergaard, "Formal specification of object-oriented modelling concepts," PhD thesis, Department of Teleinformatics, Royal Institute of Technology, Stockholm, 2000.
[10] OMG, "Unified modeling language specification" (Version 1.5), March 2003, http://www.omg.org/uml.
[11] OMG, "Semantics of a foundational subset for executable UML models" (Version 1.5 beta), September 2020.
[12] GitHub, "Foundational UML (fUML) reference implementation: an open-source implementation of the OMG foundational semantics for executable UML models (foundational UML) specification" (v1.4.4), n.d., http://modeldriven.github.io/fUML-Reference-Implementation/.
[13] M. Broy and M. Victoria Cengarle, "UML formal semantics: lessons learned," Softw Syst Model, vol.10, pp.441-446, 2011, https://doi.org/10.1007/s10270-011-0207-y.
[14] D. Calvanese, "Knowledge representation and ontologies, part 1: modeling information through ontologies," Faculty of Computer Science, European Master in Computational Logic, A.Y. 2017/2018, accessed 26-7-2020, http://www.inf.unibz.it/~calvanese/teaching/17-18-odbs/lecture-notes/KRO-3-queries.pdf.
[15] R. Vilkko, "The reception of Frege's Begriffsschrift," Hist. Math., vol.25, pp.412-422, 1998.
[16] D. D. Roberts, The Existential Graphs of Charles S. Peirce, Moulton & Co. N.V., The Hague, 1973.
[17] J. Corcoran and I. Samawi Hamid, "Schema," in The Stanford Encyclopedia of Philosophy (Fall 2016 edition), E. N. Zalta (ed.), https://plato.stanford.edu/archives/fall2016/entries/schema/.
[18] S. Al-Fedaghi, "Modeling the realization and execution of functions and functional requirements," Int. J. Comput. Sci. Inf. Secur., vol.18, no.3, pp.Page-Page, 2020.
[19] S. Al-Fedaghi and D. Al-Qemlas, "Modeling network architecture: a cloud case study," Int. J. Comput. Sci. Net. Sec., vol.20, no.3, pp.Page-Page, 2020.
[20] S. Al-Fedaghi and H. Alnasser, "Modeling network security: case study of email system," Int. J. Adv. Comput. Sci. Appl., vol.11, no.3, pp.Page-Page, 2020.
[21] S. Al-Fedaghi and M. Al-Saraf, "Thinging the robotic architectural structure," 2020 3rd International Conference on Mechatronics, Control and Robotics, Tokyo, February 22-24, 2020.
[22] S. Al-Fedaghi, "Modeling physical/digital systems: formal event-B vs. diagrammatic thinging machine," Int. J. Comput. Sci Net. Sec., vol.20, no.4, pp.208-220, 2020.
[23] S. Al-Fedaghi and E. Haidar, "Thinging-based conceptual modeling: case study of a tendering system," J. Comput. Sci., vol.16, no.4, pp.452-466, 2020, https://doi.org/10.3844/jcssp.2020.452.466.
[24] S. Al-Fedaghi and B. Behbehani, "How to document computer networks," J. Comput. Sci., vol.16, no.6, pp.423-434, 2020, https://doi.org/10.3844/jcssp.2020.723.434.

[25] S. Al-Fedaghi and J. Al-Fadhli, "Thinging-oriented modeling of unmanned aerial vehicles," Int. J. Adv. Comput. Sci. Applic., vol.11, no.5, pp.610-619, 2020, https://doi.org/10.14569/IJACSA.2020.0110575.
[26] S. Al-Fedaghi and B. Behbehani, "How to document computer networks," J. of Comput. Sci., vol.16, no.6, pp.423-434, 2020, https://doi.org/10.3844/jcssp.2020.723.434.
[27] S. Al-Fedaghi and J. Al-Fadhli, "Thinging-oriented modeling of unmanned aerial vehicles," Int. J. Adv. Comput. Sci. Applic., vol.11, no.5, pp.610-619, 2020, https://doi.org/10.14569/IJACSA.2020.0110575.
[28] S. Al-Fedaghi, "Changes, states, and events: the thread from staticity to dynamism in the conceptual modeling of systems," Int. J. Comput. Sci. Net. Sec., vol.20, no.7, pp.138-151, 2020.
[29] S. Al-Fedaghi, "Thing/machines (thimacs) applied to structural description in software engineering," Int. J. Comput. Sci. Inf. Sec., vol.17, no.8, pp.Page-Page, 2019.
[30] C. Shields, "Aristotle," The Stanford Encyclopedia of Philosophy (Fall 2020 edition), Edward N. Zalta (ed.), https://plato.stanford.edu/archives/fall2020/entries/aristotle/.
[31] J. Seibt, "Process philosophy," The Stanford Encyclopedia of Philosophy (Summer 2020 edition), Edward N. Zalta (ed.), https://plato.stanford.edu/archives/sum2020/entries/process-philosophy/.
[32] M. Heidegger, "The thing," in Poetry, Language, Thought," A. Hofstadter (transl.), Harper & Row, New York, pp.161-184, 1975.
[33] R. Klimek, "Towards deductive-based support for software development processes," in Proceedings of the 2013 Federated Conference on Computer Science and Information Systems, Kraków, pp.1377-1380, September 8-11, 2013.
[34] G. Deleuze and F. Guattari, A Thousand Plateaus: Capitalism and Schizophrenia (vol.2), B. Massumi (transl.), Athlone, London, 1988.
[35] S. M. Zamora, "Semantics," October 15, 2017, https://www.slideshare.net/SarahMaeFaithZamora/semantics-80827722.
[36] W. Hodges, "Tarski's truth definitions," in The Stanford Encyclopedia of Philosophy (Fall 2018 edition), E. N. Zalta (ed.), https://plato.stanford.edu/archives/fall2018/entries/tarski-truth/.
[37] D. Davidson, The logical form of action sentences, in N. Resher (ed.), the Logic of Decision and Action. University of Pittsburgh Press, Pittsburgh, pp.81-95, 1967. Reprinted in D. Davidson (ed.), Essays on Actions and Events, Clarendon Press, Oxford, 1980, pp.105-122.
[38] C. Maienborn, Event semantics, in C. Maienborn, K. von Heusinger and P. Portner (eds.), Semantics: An International Handbook of Natural Language Meaning (vol.1), Mouton de Gruyter, Location, pp.Page-Page. https://doi.org/10.1515/9783110589245-008.
[39] R. Nouwen, "Foundations of semantics I: truth-conditions, entailment and logic," 2011, accessed on October 25, 2020, http://www.gist.ugent.be/file/216.